\documentclass[twocolumn,aps,pre]{revtex4}

\usepackage{graphicx}
\usepackage{dcolumn}
\usepackage{amsmath}
\usepackage{latexsym}
\usepackage{verbatim}

\begin {document}

\title {Colored Sandpile}

\author{S. S. Manna\footnote[1]{subhrangshu.manna@gmail.com}}
\affiliation{B-1/16 East Enclave Housing, 02 Biswa Bangla Sarani, New Town, Kolkata 700163, India}

\begin{abstract}
      After the introduction of sandpile model a number of different variants have been studied. In most of 
   these models sand particles are indistinguishable. Here we have painted the sand particles using a 
   few distinct colors, and restrict them to move in linear trajectories only along their assigned lattice axes,
   one axis reserved for one color. Different colored particles interact among themselves through the 
   toppling of unstable sand columns. Consequently, the avalanches or in general the self-organization 
   processes in the sandpile has no overall preferred direction, though the individual particles 
   execute directed motion. For such non-abelian colored sandpiles the steady states are found to be 
   different and also the avalanche size distributions. This sandpile so defined has a non-trivial spatial 
   structure and belongs to a different universality class of sandpile models. Dynamics of a granular heap 
   with grains of different colors and properties may be described using this sandpile.
\end{abstract}

\maketitle

      The prototypical model of Self-organized Criticality (SOC) is the `sandpile', introduced by Bak, 
   Tang and Wiesenfeld (BTW) \cite {BTW}. Later, the versions with stochastic evolution rules \cite 
   {Manna} and the abelian dynamics \cite {Dhar} have also been popular. All these models establish the 
   fundamental concepts of SOC on a firm basis \cite {Tang,Dickman1,Manna2025,Fey,Dhar3,Huynh,Manna5}. 
   In general, the sandpile model had raised tremendous interest only because it generated the critical 
   behavior on its own, i.e., without fine tuning of parameters. Many different models of SOC have been 
   introduced but they are found to belong to only a few universality classes, like in the ordinary 
   critical phenomena \cite {Dhar1, Watkins, Wiese, Frontiers}. 

      In this paper, we will consider a new variant of the sandpile models, where the grains are distinguishable 
   by an additional property called color here. Such a model may be useful for studying the properties of a 
   granular heap piled with different types of grains with different properties. This model has a different behavior than previously studied 
   models, and shows a steady state that is spatially inhomogenous. The model is of interest as some features 
   of this steady state can be determined exactly, though we are not able to determine all properties of 
   interest. We have used numerical simulations to study these properties, for example the critical exponents.

      A sandpile self-organizes and responds to the external drive by diffusion of the particles 
   in the high density regions to smooth out the density differences. Such an activity is referred to as 
   an `avalanche' of particle movements which eventually terminates and the sandpile gets ready to receive 
   the next particle from the external agent. In the long-time steady state of the model, the avalanches 
   are observed to be of all length and time scales whose upper bounds are some powers of the system size. 
   Therefore, in the asymptotic limit such fluctuations of all scales is considered to be the signature of 
   the spontaneous emergence of the long-range correlation and the phenomenon is termed as the self-organized criticality. 

      There are some previous works on multi-component sandpiles in the literature. Two component abelian 
   sandpile had been studied in \cite {Alcaraz}. Self-organized criticality of the second kind had also been 
   studied in two-component sandpile \cite {Fujihara}. A sandpile of both particles and holes has been studied 
   where holes behave opposite to particles \cite {Rumani}, and a two-type internal aggregation model had been 
   studied with oil and water in \cite {Candellero}.

   {\it Definition of the model}:- The number $z$ of distinct colors is equal to the coordination number of the 
   underlying lattice. A sand particle of a particular color is assigned a specific lattice direction to move, 
   and therefore the particle trajectory is linear. For example, for the square lattice, let us name the colors 
   of the particles as 1, 2, 3, and 4 which move along the coordinate axes $+x$, $+y$, $-x$, and $-y$ respectively.

\begin{figure}[t]
\includegraphics[width=5.0cm]{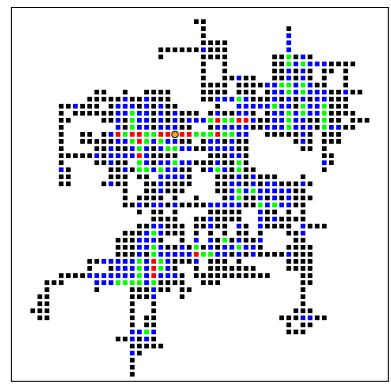}
\caption{$n_c=4$ and $L$ = 64: Different multiply toppled sites in an avalanche are marked. Sites 
	which have toppled different times are plotted using different colors e.g., 1 (black), 2 
	(blue), 3 (green), 4 (red), and 5 (orange with a circle). Their numbers are 541, 259, 71, 17, 1 respectively.
}
\label{FIG01}
\end{figure}

      The sand particles are dropped on to the square lattice of size $L \times L$ at the randomly selected 
   sites. Let the number $n(x,y)$ of sand particles sharing the lattice site $(x,y)$ at any arbitrary instant of 
   time represent the height of the sand column at that site. Such a sand column has a mixture of $n(x,y)$ 
   colored particles, and may have multiple particles of the same color. While dropping a sand particle we 
   first randomly assign a color to the particle selecting it randomly from the $z$ distinct colors with uniform 
   probability. The colors of different particles so assigned remain unchanged throughout the entire 
   dynamical evolution of the sandpile. This particle is then added at the top of the sand column at 
   the site $(x,y)$:
\begin {center}
\verb !n(x,y)! $\rightarrow$ \verb !n(x,y) + 1.!
\label {EQN01}
\end {center}
   A threshold number $n_c$ is defined for the height of stability, same for all sites. When the height 
   $n(x,y) \ge n_c$, the sand column becomes unstable and topples. In a toppling the height of the sand 
   column is reduced by $n_c$ particles:
\begin {center}
\begin {tabular}{c}
\verb !n(x,y)! $\rightarrow$ \verb !n(x,y)! - n$_{\verb !c!}$ \\
\end {tabular}
\label {EQN02}
\end {center}
   Each of $n_c$ colored particles jumps one lattice unit along the lattice axis assigned to its color. 
   Different sand columns have different mixtures of differently colored particles. Therefore, from 
   different unstable sand columns different number of sand particles jump out to different neighboring 
   sites. The $n_c$ particles are transferred from the bottom of the column using a first in first out 
   (FIFO) sequence to the neighboring sites (\verb!x!$_k$,\verb!y!$_k$) according to their colors: 
\begin {center}
\begin {tabular}{c}
\verb !n(x!$_k$, \verb!y!$_k$) $\rightarrow$ \verb!n(x!$_k$, \verb!y!$_k$) + 1.
\end {tabular}
\label {EQN03}
\end {center}
   After toppling the site $(x,y)$ is left with only the top $n(x,y) - n_c$ particles maintaining the
   same sequence.

\begin{figure}[t]
\includegraphics[width=6.0cm]{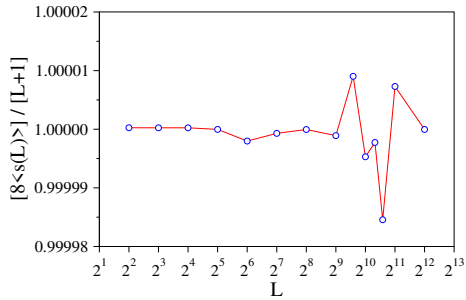}
\caption{
      For $n_c=4$: The scaled quantity $[8\langle s(L) \rangle] / [L+1]$ has been plotted against the system size
   $L$ using a semi logarithmic scale for $L$ = 4 to 4096. The entire fluctuations of the curve remained confined
   within $1 \pm 0.00002$ which confirms the relation $\langle s(L) \rangle = (L+1)/8$.
}
\label{FIG02}
\end{figure}

      The heights of the sand columns of a stable configuration are less than the threshold height: 
   $0 \le n(x,y) < n_c$ at all sites $(x,y)$. In addition, the colors of the sand particles in each column are random. 
   The height profile of the entire sandpile is determined by the random choice of the colors of the 
   particles at the very initial stage when they were dropped and their randomly selected drop locations. Once a
   particle is dropped, the entire evolution of the avalanche so triggered is completely deterministic
   similar to the BTW model. In previous sandpile models if a sand particle is tagged it selects one of the neighboring
   sites randomly during a toppling, and therefore their paths are random walks. In contrast, here the trajectories
   of particles are straight lines and therefore the colored sandpile belongs to a different universality class.
   It may be noticed that the BTW model is not retrieved if we use all particles of the same color (see Supp01).

      Since the sand particles are colored and distinguishable among different colors, the sequence
   in which a lattice site receives particles is important. Let us consider two unstable sites $i$ and 
   $j$ which have one common neighbor site $k$. Since the colors of particles received by $k$ from $i$ and $j$ are
   different, the sequence of particles of the sand column at $k$ would depend if the site $i$ is 
   toppled first and then the site $j$ or the sites were toppled in the reverse sequence. This implies 
   that the color distribution of particles in the stable state depends on the sequence in which different 
   unstable sites have been updated and the rule which particles follow to move out in a toppling event. Therefore, 
   the colored sandpile follows the non-abelian sandpile dynamics in comparison to the BTW sandpile being 
   the abelian sandpile \cite {Dhar}.

\begin{figure}[t]
\includegraphics[width=6.0cm]{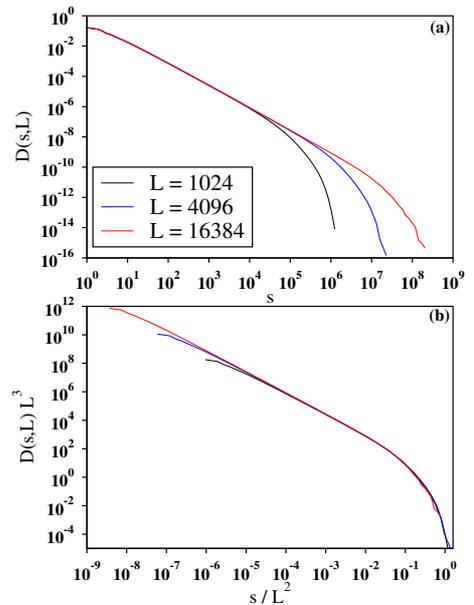}
\caption{
   For $n_c=4$: 
   (a) Probability distribution $D(s,L)$ of the avalanche sizes have been plotted against the
   avalanche size $s$ for three system sizes $L$.
   (b) The same data have been re-scaled and a plot of $D(s,L)L^{3}$ against $s / L^2$ exhibits
   a nice collapse of the data.
}
\label{FIG03}
\end{figure}

      A lattice site can topple multiple times within the same avalanche. In the BTW sandpile lower 
   multiply toppled zones surround the higher multiply toppled zones in general and form compact clusters
   \cite {Grass}. Such internal structures of the avalanches are absent in colored sandpile. In Fig.
   \ref {FIG01} we have shown the picture of an avalanche that toppled a total number of $s$ = 1345 
   times covering 889 distinct sites on a lattice of size $L =64$ where a single site toppled the maximum 
   of five times. It may be noticed that the avalanche is not compact like BTW sandpile, different
   clusters of sites which have toppled the same number of times are not connected clusters.

      However, the most important difference is the avalanche cluster of Fig. \ref {FIG01} has propagated in all directions
   equally well, though each particle moves along a specified lattice direction determined by its color.
   This happens only due to the fact that the colored sandpile is a random mixture of particles of 
   all $z$ colors. Here, the colors of the particles have been chosen with equal probabilities which
   ensures that the branching process is isotropic. Compared to the directed sandpiles \cite 
   {Ram,Pac} the avalanche clusters have no overall preferred direction (see Supp01).
   Note that the model has reflection symmetry about the $x$ and $y$ axes, with a corresponding 
   change in the color labels $1<->3$, $2<->4$. If the colors of the particles are not assigned with equal probability, 
   then the avalanche clusters will be anisotropic.

\begin{figure}[t]
(a) \\
\includegraphics[width=6.5cm]{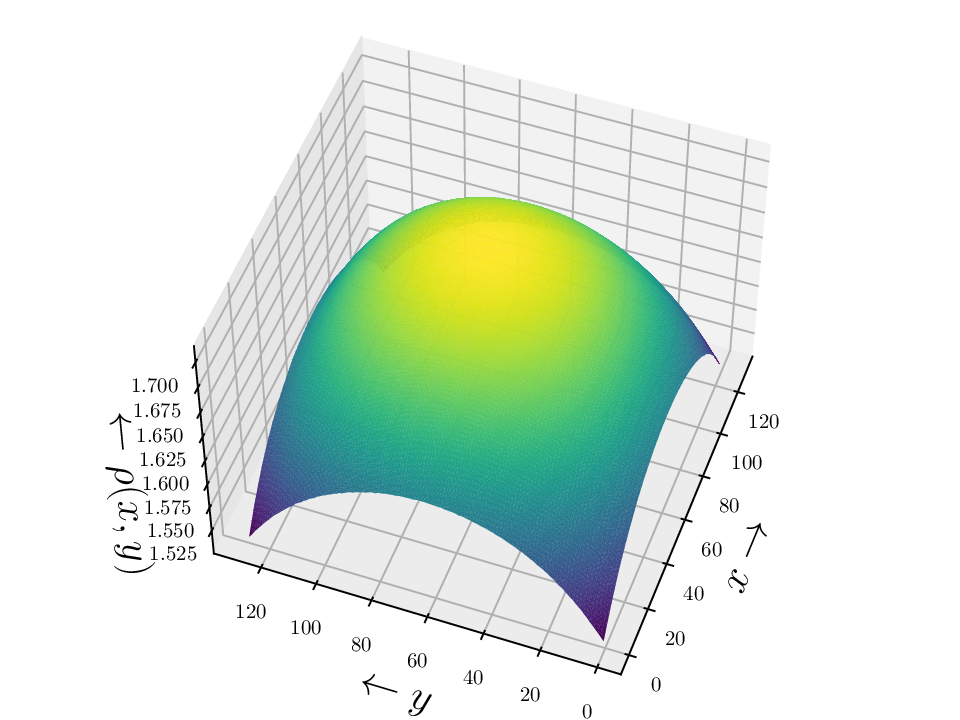} \\
(b) \\
\includegraphics[width=6.5cm]{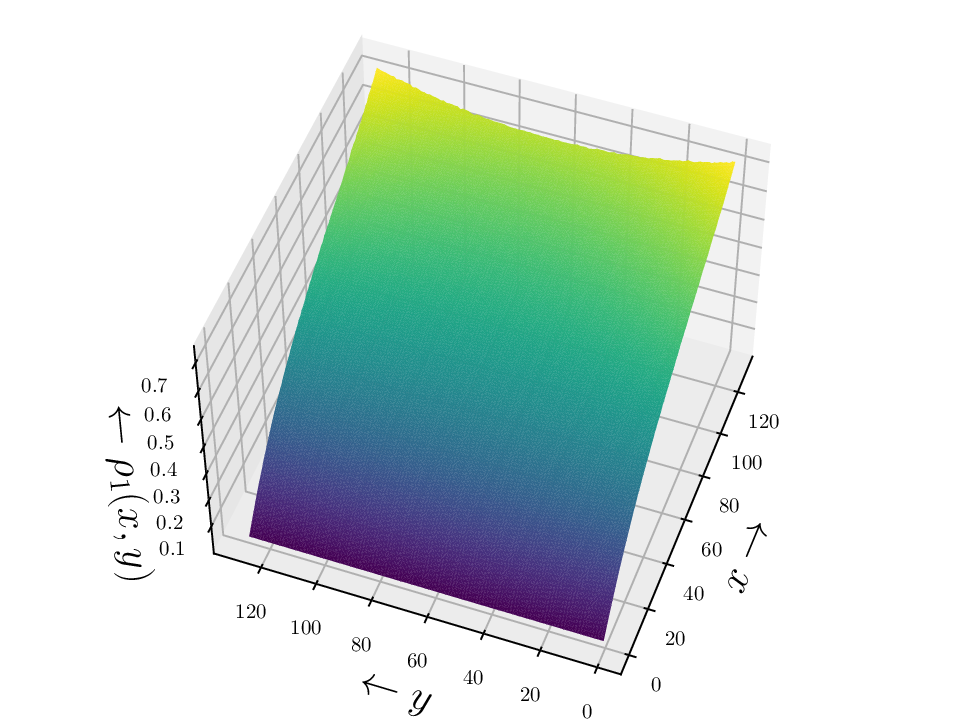}
\caption{Surface plots of the average particle density profile of the colored sandpile for $n_c=4$ and $L=128$:
	(a) $\rho(x,y)$ profile of all four colored particles and,
	(b) $\rho_1(x,y)$ profile of only the color 1 particles.
}
\label{FIG04}
\end{figure}

      From the definition of the model, some properties of the steady state can be deduced quite easily. 
   The size $s$ of an avalanche is measured by the total number of topplings in the avalanche. In the 
   stationary state, the average size of the avalanches is denoted by $\langle s(L) \rangle$ for a sandpile 
   of size $L$.  A particle dropped at random needs $(L+1)/2$ topplings  on the average to get out. In the 
   case of $n_c=4$, toppling of a sand column creates four particle jumps. Therefore in the stationary state, 
   the average size of an avalanche is
\begin{equation}
\langle s(L) \rangle = (L+1)/8.
\label {EQN01}
\end{equation}
   Numerically the average avalanche size is estimated by simulating a large number of avalanches and by 
   taking a simple average of their sizes. In Fig. \ref {FIG02} we have plotted $[8\langle s(L) \rangle]/[L+1]$ 
   against $L$. For the entire range of $L$ the vertical coordinate varied within the range $1 \pm 0.00002$ 
   which confirms Eqn. \ref {EQN01}.

\begin{figure}[t]
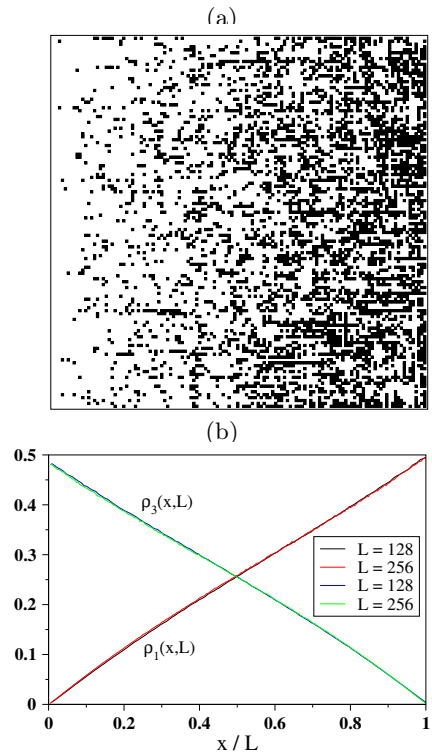

(a) \\
\hspace*{0.45cm}\includegraphics[width=5.0cm]{figure05a.eps} \\
(b) \\
\includegraphics[width=5.5cm]{figure05b.eps}
\caption{For $n_c=2$ and $L$ = 128: 
   (a) Locations of only the color 1 particles in a stable state have been plotted. 
   (b) The density $\rho_1(x,L)$ of color 1 particles has been plotted against $x / L$ 
   for $L = 128$ (black) and 256 (red). Similarly, the density $\rho_3(x,L)$ of color 3 
   particles have been plotted using blue and green colors respectively.
}
\label{FIG05}
\end{figure}

      Let $j_{\kappa}(x,y)$ denote the average number of particles leaving the site $(x,y)$ per added particle. Then, clearly,
\begin{equation}
j_{\kappa}(x+ \hat{e}_{\kappa},y) = j_{\kappa}(x,y) + 1/L^2,
\label {EQN02}
\end{equation}
where $\hat{e}_{\kappa}$ is a unit vector in the direction of movement of grains of color $\kappa$. This equation, say for 
$\kappa=1$, with the boundary condition $j_{1}(x,y)=0$, at $x=0$ is solved to give
\begin{equation}
j_1(x,y) = x/L^2,
\label {EQN03}
\end{equation}
and similar equations for other components. Now, consider the site $(x,y)$. The average number of particles exiting this site per added particle, in the directions $1,2,3,4$ are in the ratio
$x : y : (L-x+1) : (L-y+1)$. The toppling condition does not depend on color, it seems plausible that in the steady state, the 
average number of particles of color $1, 2, 3, 4$ at the site $(x,y)$ would also be in the same ratio. This expectation is 
satisfied approximately, but our data reveals significant deviations from this prediction in the steady state, due to 
correlations in the colors of different topplings in an avalanche. The average density of particles $\rho(x,y)$ has a 
non-trivial space-dependent profile, and the fractional number of particles of colors $1$ to $4$ at site $(x,y)$ will be  
nearly proportional to $x, y, (L-x+1), (L-y+1)$. 

\begin{figure}[t]
\includegraphics[width=5.5cm]{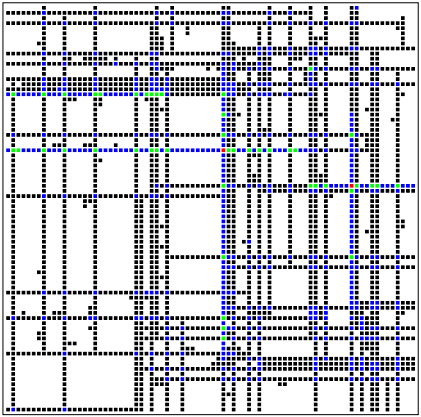} \\
\caption{Plot of the toppled sites of an avalanche of the flush all version of the colored sandpile
	on $L$ = 80 square lattice. The number of sites which have toppled 1 (black), 2 (blue), 3 (green) 
	and 4 (red) times are 2328, 472, 45, 2 respectively.}
\label{FIG06}
\end{figure}
   
      The probability distributions $D(s,L)$ of the avalanche sizes have also been measured for the 
   colored sandpile for the system sizes $L$ = 1024, 4096 and 16384 and have been plotted in Fig. \ref 
   {FIG03}(a). In Fig. \ref {FIG03}(b) we scale the axes and re-plot the same data as $D(s,L)L^3$ 
   against $s/L^2$. A good data collapse supports the finite-size scaling form,
\begin {equation}
D(s,L) L^{\beta} \sim {\cal G}[s / L^{\alpha}],
\label {EQN04}
\end {equation}
   where, ${\cal G}(x)$ is an universal scaling function that is expected to decay in a power law 
   form $x^{-\tau}$ in the limit of $x \to 0$ i.e., in the asymptotic limit of $L \to \infty$. The 
   scaling exponents are: $\beta=3$, $\alpha=2$ and $D(s) \sim s^{-\tau}$ implies $\tau = \beta / \alpha = 1.5$.
   Further, to check if the values of the exponents depend on the local regions where the avalanches
   are triggered, we have divided the entire lattice into nine equal square regions. Three different 
   probability distributions have been calculated whose triggering points are in the (i) four corner 
   boxes, (ii) four boxes at the middle of the sides, and (iii) at the central box. Again from the 
   scaling analysis we find the values of exponents $\alpha$, $\beta$ and $\tau$ are very close to 
   those reported above (see Supp02).

      The life-time $T$ of an avalanche has been estimated by the total number of synchronous updates 
   of the system needed to become stable. At any arbitrary intermediate time $T$ all the unstable sites are 
   ordered in a random sequence and then simultaneously toppled. From every toppling site, $n_c$ particles 
   simultaneously jump to their neighboring sites as per their intrinsic colors. The neighboring sites whose
   heights are more than $n_c-1$ constitute the list of unstable sites in time $T+1$ (see Supp03).

      In a similar scaling analysis of the probability distribution $D(T,L)$ for the life-times, a plot of
   $D(T,L)L^{1.84}$ against $T/L$ exhibits nice collapse of the data. Therefore, in the asymptotic limit of 
   $L \to \infty$ the power law form is $D(T) \sim T^{-\tau_T}$ with $\tau_T = 1.84$. Another attempt by 
   plotting $D(T,L)L^{2}$ against $T/L$ exhibits a worse data collapse and therefore $\tau_T = 2$ is ruled 
   out (see Supp04).

      The average value of the life-times $\langle T(L) \rangle$ exhibits a good linear fit against $L^{0.1474}$
   between $L$ = 64 to 16384. Therefore, we conclude a variation $\langle T(L) \rangle \sim L^{0.1474}$ for large $L$
   (see Supp05).

      How the particles are distributed on the lattice in the stationary state? Our numerical study reveals
   the average density $\rho(x,y)$ of particles of all colors is certainly not uniform at all. The effect of 
   the open boundary is clearly present in the density profile. The density is maximum at the centre of the 
   lattice which gradually decreases to all four sites. Further, its value is more at the middle points of 
   the sides compared to the four corner points. A surface plot of $\rho(x,y)$ has been shown in Fig. \ref 
   {FIG04}(a) where a large number of uncorrelated configurations in the stationary state have been averaged 
   to obtain $\rho(x,y)$. 

      We have first estimated the overall average density $\rho(L)$ for different values of $n_c$ and system
   sizes. Because of the presence of the open boundary and for the finite size of the lattice, the average 
   particle density $\rho(L)$ is $L$ dependent. As the system size increases, the average density $\rho(L)$ 
   gradually approaches to its asymptotic value $\rho_c$ following finite size correction
\begin {equation}
\rho(L) = \rho_c + A. L^{-1/\nu}.
\label {EQN05}
\end {equation}
   Estimates of $\rho(L)$ for sandpile sizes $L$ = 512, 1024, 2048, and 4096 are plotted against $L^{-1/\nu}$. 
   Tuning the values of $1/\nu$ we obtained their best fitted values and the corresponding $\rho_c$ values which
   grow with $n_c$ from 0.5536 for $n_c=2$ to 2.1856 for $n_c=5$ and have been listed in TABLE I.

\begin{table*}[t]
\centering
	\caption{Estimates of $\rho_c$, $f_i$ and $1/\nu$ for different values of $n_c$}
\begin{tabular}{|c|cc|cc|cc|cc|cc|cc|} \hline
$n_c$ & $\rho_c$ & $1/\nu$ & $f_0$    &$1/\nu_0$& $f_1$  &$1/\nu_1$& $f_2$  & $1/\nu_2$& $f_3$  &$1/\nu_3$& $f_4$  &$1/\nu_4$\\ \hline
2     & 0.5536   & 0.503   & 0.4464   & 0.503   & 0.5536 & 0.503   &        &          &        &         &        &        \\
3     & 1.0999   & 0.261   & 0.2765   & 0.941   & 0.3481 & 0.772   & 0.3757 & 0.439    &        &         &        &        \\
4     & 1.6426   & 1.767   & 0.1973   & 0.640   & 0.2460 & 0.593   & 0.2729 & 1.698    & 0.2838 & 0.420   &        &        \\ 
5     & 2.1856   & 1.036   & 0.1526   & 0.758   & 0.1873 & 0.639   & 0.2099 & 0.707    & 0.2224 & 0.900   & 0.2279 & 0.585 \\ \hline
\end{tabular}
\end{table*}

      In a more detailed analysis, the fractions $f_k$ of sites occupied with $k$ particles of any color 
   are also estimated. For example, for $n_c=2$, the lattice sites are either occupied by single particle 
   or they remain empty. Simultaneously, we have also estimated the fraction of sites $f_k$ occupied by 
   the $k = 0, ... , n_c-1$ particles for the same lattice sizes. They are extrapolated using Eqn. \ref 
   {EQN01} for $n_c$ = 3, 4, and 5 and their asymptotic values are listed in Table 1 together with the
   $1/\nu$ values.

      The linear trajectories of particles of a specific color determine their density profile. A color 1 
   particle, dropped at a randomly selected site, moves certain distance along the $+x$ axis within an 
   avalanche and stops there. Thus, a density gradient is generated (Fig. \ref {FIG04}(b)). The density 
   is minimum at the left boundary $(x=1)$ and is maximum at the right boundary $(x=L)$ (Fig. \ref {FIG05}(a)). 
   In Fig. \ref {FIG05}(b) $\rho_1(x,L)$ values have been plotted against $x$ which grow linearly
   with $x$ for $L$ = 128, and 256 which almost overlap. On the same plot the density $\rho_3(x,L)$ of 
   color 3 particles decreases linearly to the right boundary.


{\it Colored sandpile in one dimension}:- The colored sandpile is non-trivial in one dimension as well.
   There are only two colors, $+x$ and $-x$ and we used $n_c=2$. Again, a similar scaling analysis of 
   $D(s,L)$ data gives $\beta=2.0$ and $\alpha=1.5$, implying $\tau=4/3$. For life-time distribution 
   $D(T,L)$ the similar scaling exponents are, $\beta_T=1.5$ and $\alpha_T=1$, giving $\tau_T=1.5$. 
   The profile of the average particle density $\rho(x,L)$ has a maximum at the center of the lattice 
   that fits approximately an inverted parabola (see Supp06).

{\it Flush all version of the colored sandpile}:-
      In a toppling all the particles of the unstable sand column jump to their neighboring sites 
   according to their intrinsic colors. This is referred as the `flush all' version of the colored 
   sandpile. Similar flush all rule was used in the original Manna sandpile \cite {Manna}. For $n_c=2$,
   a sand column becomes unstable only when two or more particles share the same lattice site at the 
   same time. The crucial difference in this version is, whenever two particles of the same color 
   meet at one site they form a tightly bound pair and then on make steps together only along the
   same lattice axis. Irrespective of the size $L$ of the lattice they move together all the way to 
   the boundary and then finally drop out from the sandpile. Moreover, some other particles of the 
   same color who were occupying sites on the same path would also join in and this way the single
   site toppling front gradually grows in size and finally jumps out of the system at the boundary.
   During the progress of this avalanche, several other branches will also be created by particles 
   of other colors, which would also terminate at the lattice boundary. 

      A typical picture of an avalanche in the flush all model has been displayed in Fig. \ref {FIG06}. The sites
   which have toppled have been plotted by dots: black dots represent singly toppled sites, where as the
   red dots represent doubly toppled sites. The black dots form black straight lines which criss-cross themselves and the 
   red dots are situated almost always at the crossing of these black lines. The green dot marks the site
   where the sand particle had been added.

       This picture shows that avalanches of this model flush out large amounts of sand from the system.
   Even then there exists a distinct stationary state for every system size $L$ that balances the inflow 
   rate and the average outflow rate. The average density $\rho(L)$ of sand particles in the stationary 
   state decreases linearly as $L^{-0.3655}$. On extrapolation to the limit of $L \to \infty$ the 
   asymptotic value of particle density nearly misses the origin of the coordinate axes (see Supp07).

      In a scaling analysis of the avalanche size distribution $D(s,L)L^{2.55}$ plotted against $s / L^{1.7}$
   gives the best collapse of the data for the regime of large avalanche sizes implying $\tau = 2.55 / 1.7 = 1.5$. 
   For the small avalanche sizes the distribution has flat portions which can be scaled by plotting 
   $D(s,L)L^{1.105}$ against $s / L$ (see Supp08).

      To summarize, a sandpile model with colored sand grains is introduced and studied. This model has an 
   interesting non-trivial steady state in which the density of grains of each color shows a roughly constant 
   gradient, but the sum of the densities of all colors is nearly uniform. The sandpile is non-abelian, and 
   shows exponents different from other sandpiles models, and we have estimated these by simulations. It is 
   hoped that further studies would help understand the unusual spatial structures in the critical state of 
   the model.


      Many helpful discussions with Prof. D. Dhar are thankfully acknowledged. I also thank Dr. Sumanta 
   Kundu and Mr. Animesh Hazra for running some big jobs for me.


\begin{thebibliography}{90}

\bibitem {BTW} P. Bak, C. Tang and K. Wiesenfeld, Self-organized criticality: An explanation of $1/f$ noise, 
	 Phys. Rev. Lett., {\bf 59}, 381 (1987), https://doi.org/10.1103/PhysRevLett.59.381.
\bibitem {Manna} S. S. Manna, Two-state model of self-organized criticality, 
	 J. Phys. A: Math. Gen.,  {\bf 24} , L363 (1991), DOI:10.1088/0305-4470/24/7/009.
\bibitem {Dhar} D. Dhar, Self-Organized Critical State of Sandpile Automation Models, 
	 Phys. Rev. Lett. {\bf 64}, 2837 (1990), https://doi.org/10.1103/PhysRevLett.64.1613.
\bibitem {Tang} C. Tang and P. Bak, Critical Exponents and Scaling Relations for Self-Organized Critical Phenomena, 
	 Phys. Rev. Lett. {\bf 60}, 2347 (1988), https://doi.org/10.1103/PhysRevLett.60.2347.
\bibitem {Dickman1} R. Dickman, A. Vespignani, and S. Zapperi, Self-organized criticality as an absorbing-state phase transition,
	 Phys. Rev. E {\bf 57}, 5095 (1998), https://doi.org/10.1103/PhysRevE.57.5095.
\bibitem {Manna2025} S. S. Manna, Describing self-organized criticality as a continuous phase transition,
         Phys. Rev. E {\bf 111}, 024111 (2025), DOI: 10.1103/PhysRevE.111.024111.
\bibitem {Fey} A. Fey, L. Levine, and D. B. Wilson, Driving Sandpiles to Criticality and Beyond,
	 Phys. Rev. Lett. 104, 145703 (2010), DOI:https://doi.org/10.1103/PhysRevLett.104.145703.
\bibitem {Dhar3} D. Dhar, Some results and a conjecture for Manna's stochastic sandpile model,
	 Physica A {\bf 270}, 69 (1999), https://doi.org/10.1016/S0378-4371(99)00149-1.
\bibitem {Huynh} H. N. Huynh, G. Pruessner, and L. Y. Chew, The Abelian Manna model on various lattices in one and two dimensions,
	 J. Stat. Mech., P09024 (2011), DOI 10.1088/1742-5468/2011/09/P09024.
\bibitem {Manna5} S. S. Manna, Nonstationary but quasisteady states in self-organized Criticality, 
	 Phys. Rev. E. {\bf 107}, 044113 (2023), https://doi.org/10.1103/PhysRevE.107.044113.
\bibitem {Dhar1} D. Dhar, The Abelian sandpile and related models,
	 Physica A {\bf 263}, 4 (1999), https://doi.org/10.1016/S0378-4371(98)00493-2.
\bibitem {Watkins} N. W. Watkins, G. Pruessner, S. C. Chapman, N. B. Crosby, and H. J. Jensen, 
	 25 Years of Self-organized Criticality: Concepts and Controversies, 
	 Space Sci Rev, {\bf 198}, 3 (2016), https://doi.org/10.1007/s11214-015-0155-x.
\bibitem {Wiese} K. J. Wiese, Theory and experiments for disordered elastic manifolds, depinning, avalanches, and sandpiles,
	 Rep. Prog. Phys. {\bf 85}, 086502 (2022), https://doi.org/10.1088/1361-6633/ac4648.
\bibitem {Frontiers} S. S. Manna, A. L. Stella, P. Grassberger, and R. Dickman, Self-organized Criticality, Three Decades Later,
         Frontiers in Physics, 2022, DOI 10.3389/978-2-88974-219-6.
\bibitem {Alcaraz} F. C. Alcaraz, P. Pyatov, and V. Rittenberg, Two-component Abelian sandpile models,
	 Phys. Rev. E, {\bf 79}, 042102 (2009); 10.1103/PhysRevE.79.042102.
\bibitem {Fujihara} A. Fujihara, T. Ohtsuki, and T. Nakagawa, Two-component sandpile model: self-organized criticality of the second kind
	 arXiv:cond-mat/0506759; 10.48550/arXiv.cond-mat/0506759
\bibitem {Rumani} R. Karmakar and S. S. Manna, Particle–hole symmetry in a sandpile model
	 J. Stat. Mech. (2005) L01002; 10.1088/1742-5468/2005/01/L01002.
\bibitem {Candellero} E. Candellero, S. Ganguly, C. Hoffman, L. Levine,
	 Oil and water: A two-type internal aggregation model,
	 Ann. Probab. {\bf 45}, 4019 (2017); 10.1214/16-AOP1157.
\bibitem {Grass} P. Grassberger and S. S. Manna, Some more sandpiles, 
	 J. Phys. France {\bf 51}, 1077 (1990), DOI: 10.1051/jphys:0199000510110107700.
\bibitem {Ram} D. Dhar and R. Ramaswamy, Exactly solved model of self-organized critical phenomena,
	 Phys. Rev. Lett. {\bf 63}, 1659 (1989); https://doi.org/10.1103/PhysRevLett.63.1659.
\bibitem {Pac} M. Paczuski and K. E. Bassler, 
	 Theoretical results for sandpile models of self-organized criticality with multiple topplings, 
	 Phys. Rev. E. {\bf 62}, 5347 (2000); doi.org/10.1103/PhysRevE.62.5347.
\end {thebibliography}
\end {document}


\vskip 0.2 cm
\centerline {\bf {\small Supplementary Materials}}
\vskip 0.2 cm

\small {
{\centerline {\bf Colored Sandpile}}
{\centerline {\bf S. S. Manna}}

\begin {itemize}

\item [01.] {\bf Why the universality class of the colored sandpile is different from that of the directed sandpile?}

	(a) The avalanche clusters in the colored sandpile are not anisotropic like the directed sandpile.
	It is clear from the picture of the avalanche cluster in FIG. 1 of the manuscript that the cluster 
	has no perpendicular time direction and space direction like the avalanche clusters in the directed 
	sandpile. As per the definition of the model, every colored particle moves in a linear trajectory.
	However, since the colored sandpile is a random mixture of particles of all colors 
	in equal proportions, dropping a single particle of a particular color excites through the topplings
	particles of colors and they start moving along all lattice axes. Therefore, the avalanche of particle 
	movements grow and propagate in the medium isotropically along all lattice axes. This is the fundamental 
	difference between the dynamics of the colored sandpile and the directed 
	sandpile and it is the main reason why the colored sandpile belongs to a different universality class.

	(b) If we paint all particles using the same color 1 then the sandpile would behave like the following.
        Let the critical threshold number $n_c$ for toppling be equal to 4. Therefore in a toppling, 4 particles
	would jump out of the toppling site, but all would land up at the same neighboring site along the $+x$
	direction. When a particle is added at the site $(x,y)$, if an avalanche is triggered the avalanche would
	propagate only along the $+x$ direction and continue to move all the way till it reaches the boundary
	at $x = L$. There is no branching of the single site toppling front. Therefore, the entire lattice is
	effectively divided into $L$ such one dimensional systems of length $L$ along $L$ parallel lattice axes.
	The avalanches on different axes, or even on neighboring axes do not interact at all, i.e., completely
	independent. The avalanche size is exactly equal to $(L-x+1)$ and these avalanches occur with equal 
	probabilities. In each avalanche exactly 4 particles leave the system. Therefore, such a model is indeed a
	directed model, but it has no similarity with the `Directed Sandpile Model' of Dhar and Ramaswamy [20].

	(c) Here we compare the colored sandpile with the directed sandpile model of Paczuski and Bassler [21]. In
	their model in the stationary state, all possible stable configurations appear. With $n_c = 4$, such
	states are $4^{L^2}$ in number. Paczuski and Bassler suggested that unlike the BTW model, all stable
	states appear in the stationary states of their directed sandpile, and they are equally likely, i.e., 
	they occur with equal probability.

	   We numerically compare our colored sandpile with Paczuski and Bassler model at this point. We consider a $2 \times 2$
	lattice and followed the evolution of the colored sandpile on this substrate till $10^{10}$ particles are
	dropped. In the stable configurations of the stationary state the number of particles $n(x,y)$ at a site varies
	in the range $\{0,3\}$. Therefore one can associate a number between 0 and 255 calculated
	by: $k=64.n(2,2)+16.n(2,1)+4.n(1,1)+1.n(1,2)$ with every height configuration.
	In the stationary state we have calculated the probability distribution $P(k)$ for occurrence of these states.
	All $4^4$ = 256 states appeared, similar to Paczuski and Bassler model but their probabilities are not uniform
	as shown in the FIG. 1. This is another primary difference that the stable configurations of the
        stationary states of the colored sandpile are not equally likely like Paczuski and Bassler directed sandpile.
\begin{figure}[h]
\includegraphics[width=12.0cm]{SuppFigure01.eps} \\
\caption {Probability distribution of 256 different stable states of the $2 \times 2$ colored sandpile.}
\label{FIG01}
\end {figure}

\item [02.] {\bf Statistics of avalanches triggered at different regions of the sandpile}

      An important feature of the colored sandpile is that the steady state is spatially inhomogeneous. Then, 
   unlike the BTW or the directed ASM, the typical size of avalanches of a given color particle dropped randomly 
   on the lattice depends strongly on where the grain is dropped. To test if the avalanches triggered at different 
   local regions of the sandpile follow different statistics we have divided the entire $L \times L$ lattice into 
   nine equal square boxes. Three different sets have been
	   formed: the corner boxes are numbered
	   as 1, 3, 7 and 9; the four boxes at the middle of the sides are numbered as 2, 4, 6, and 8; where as the box
	   at the centre is numbered as 5. The sandpile dynamics have been simulated as described in the model where
	   sand particles have been dropped at randomly selected lattice sites with uniform probability, assigning
	   randomly selected colors. The avalanche size distribution data $D(s,L)$ have been collected separately for the
	   three sets. The sizes of the avalanches which originated at the corner boxes have been collected in the
	   first distribution and $D(s,L)$ vs. $s$ data for three system sizes have been plotted in the left panel of FIG. 2.
	   Similarly the data for the middle side boxes have been plotted in the middle panel. Finally, the data for
	   the central box have been plotted in the right panel. We observe that the data collapse turn out to be the
	   best in all three cases when the finite size scaling exponents assume the values $\beta$ = 3 and $\alpha$ = 2,
	   leading to $\tau$ = 3/2.
	   This exercise indicates that the avalanches triggered anywhere in the lattice are most likely to be
	   equivalent and obey the same statistics of their probability distribution of sizes.

\begin{figure}[t]
\includegraphics[width=5.5cm]{SuppFigure02-1.eps} \includegraphics[width=5.5cm]{SuppFigure02-2.eps} \includegraphics[width=5.5cm]{SuppFigure02-1.eps} \\
\caption{Three different sets of probability distributions of avalanche sizes $D(s,L)$ against $s$ for the
	avalanches which originated at the three distinct regions of the sandpile.}
\label{FIG02}
\end {figure}

\item [03.] {\bf Definition of one intra-avalanche time step}

	An `avalanche' is the evolution process of the sandpile in response to the addition of a single
        sand particle. This process takes place in succession, one set of unstable sites topple to create
	a new set of unstable sites. More specifically, the intra-avalanche time is denoted by $T$. An
	arbitrary unit of time evolution from $T$ to $T+1$ consists of the following three intermediate steps:

	(i) A list is made of all lattice sites which are unstable at time $T$. The entries of the sites
	in this list are then randomised with uniform probability to avoid any possible bias due to
	non-abelian property of the sandpile dynamics.

	(ii) The microstate of the entire sandpile at any arbitrary intermediate time is specified by the
	colors of all the sand particles at all lattice sites. This data have been stored using a 3-$d$
	array $s(x,y,n(x,y))$. It means there are $n(x,y)$ particles residing at the lattice site $(x,y)$
	at that time. E.g., $n(x,y) = 1$ means there is only one particle at $(x,y)$ and its color is
	$s(x,y,1)$. Similarly, $n(x,y) = 5$ means there are five particles at $(x,y)$ and the fifth
	particle has the color $s(x,y,5)$, and so on. Therefore, $s(x,y,1)$ is the color of the particle
	at the bottom of the column, where as $s(x,y,n(x,y))$ refers the color of the most recent particle
	added to the top of the column at $(x,y)$ counted from the bottom.

	During the second intermediate step all unstable sites are toppled one by one as per the random
	sequence made in (i). To execute the FIFO rule, for each site $n_c$ particles from the bottom of
	the column are transferred to the neighbouring sites as per their individual colors. Since the
	colors of the particles in the stack are random, therefore different neighboring sites receive
	different number of particles in general. The height $n(x,y)$ of the sand column at site $(x,y)$
	is reduced by $n_c$ and $n(x,y)-n_c$ particles come down and fill same number of sites from the
	bottom of the sand column. It may happen quite naturally that a site receives multiple particles
	from one or more neighbors.

        (iii) In the third step, each site of the random list and their neighbors are searched to find
	out the unstable sites with $n(x,y) \ge n_c$ and their list is made for the time $T+1$.

	These three steps together constitute a unit of intra-avalanche time.

\item [04.] {\bf Probability distribution $D(T,L)$ of avalanche life-times $T$ of colored sandpile.}
\begin{figure}[h]
\includegraphics[width=17.0cm]{SuppFigure03.eps}
\caption{Plots of the probability distribution $D(T,L)$ of the life-times $T$ of the avalanches in colored sandpile
and their finite size scaling.}
\label{FIG03}
\end{figure}

      The life-time $T$ of an avalanche has been estimated by the number of synchronous updates required
   for the system to become stable. Different avalanches have different life-times. One calculates the
   probability distribution $D(T,L)$ of the avalanche life-times collecting the data when the system is
   settled in the stationary state. In FIG. 3(a) of the supplementary section we have plotted the
   distributions $D(T,L)$ against $T$ for three widely separated system sizes $L$ = 1024, 4096, and 16384. A finite size
   scaling analysis of the same data have been performed in FIG. 3(b) where, $D(T,L)L^{1.84}$ have been
   plotted against $T / L$. The data collapse implied $\beta_T = 1.84$ and $\alpha_T=1$ and therefore the
   life-time distribution exponent $\tau_T$ = 1.84/1=1.84. To check if $\beta_T = 2$ can also be a possibility, the
   finite-size scaling has also been performed using $\beta_T = 2$. Here in FIG. 3(c) $D(T,L)L^2$ have been
   plotted against $T/L$ again for the same system sizes. It is observed that the three curves separate out
   from one another and the data collapse is not at all a good one.

\item [05.] {\bf Variation of the average life-times $\langle T(L) \rangle$ with system size $L$.}
\begin{figure}[h]
\includegraphics[width=6.0cm]{SuppFigure04.eps}
\caption{The average life-time $\langle T(L) \rangle$ have been plotted for the nine different sandpiles
	of size varying from $L$ = 64 to 16384.}
\label{FIG04}
\end{figure}

	The life-time $T$ defined above have been estimated for every avalanche when the sandpile is settled
   in the stationary state. These values of $T$ vary widely but their average $\langle T(L) \rangle$ assumed
   specific system size $L$ dependent values. Accurate values have been obtained when very large sample sizes 
   have been considered for averaging. In the FIG. 4 we have plotted $\langle T(L) \rangle$ against 
   $L^{0.1474}$ so that the curve fits to an excellent straight line in the range for $L$ = 64 to 16384. 
   The equation of the best fitted straight line is: $\langle T(L) \rangle = -8.6004 + 6.1373.L^{0.1474}$.

\item [06.] {\bf Probability distributions of avalanche size $D(s,L)$ and life-times $D(T,L)$ for the colored sandpiles in
	one dimension.}
\begin{figure}[t]
\includegraphics[width=5.5cm]{SuppFigure05-1.eps} \includegraphics[width=5.5cm]{SuppFigure05-2.eps} \includegraphics[width=5.5cm]{SuppFigure05-3.eps} \\
\caption{Plots of the avalanche size distributions for the colored sandpile in one dimension.}
\label{FIG05}
\end {figure}

      The colored sandpile in one dimension. The probability distribution $D(s,L)$ for the avalanche sizes
   have been plotted in column (a) of FIG. 5 against the avalanche size $s$ for three sandpiles of size
   $L$ = 1024, 4096 and 16384. Their finite size scaling $D(s,L)L^2$ against $s/L^{1.5}$ for the data
   collapse has also been shown, implying the values of the scaling exponents $\beta_s=2$ and $\alpha_s=1.5$ and
   the avalanche size exponent $\tau_s=4/3$. Secondly, the life-time distribution $D(T,L)$ and their
   scaling have been shown in column (b) of FIG. 5. Here, the values of the scaling exponents $\beta_T=2$ and $\alpha_T=1.5$ and
   the avalanche life-time exponent $\tau_T=4/3$. In the last column (c) of FIG. 5 the values of the average
   avalanche size $\langle s(T,L) \rangle$ have been plotted against the life-time $T$. In the
   intermediate regime, the variation follows a power law: $\langle s(T,L) \rangle \sim T^{1.4899}$.

\item [07.] {\bf Flush all version: Variation of the average density of colored particles.}

\begin{figure}[h]
\includegraphics[width=17.0cm]{SuppFigure06.eps}
	\caption{Flush all colored sandpile: 
	(a) Variation of the density $\rho_1(x,L)$ of only the color 1 (black line) particles and the density $\rho_3(x,L)$ 
	of only the color 3 (red line) particles in the stationary state as a function of $x$ 
	coordinates for two different system sizes, $L$ = 256 (upper) and 1024 (lower).
	(b) The total density $\rho_1(x,L) + \rho_3(x,L)$ of the color 1 and color 3 particles in the stationary state as a function of $x$.
	(c) The average overall site density $\rho(L)$ of particles of all colors have been plotted against $L^{-0.3655}$ to obtain 
	the best straight line which is the extrapolated to the limit of $L \to \infty$. The best least square fitted straight line 
	has the equation: $\rho(L) = -0.031703 + 1.3806 L^{-0.3655}$.}
\label{FIG06}
\end{figure}

        In the flush all version of the colored sandpile all sand particles of the unstable sand column are flushed out
	to their neighboring sites as per their individual colors. This is the only difference. Here the threshold $n_c$ for the 
	stability of a sand column is assumed to be 2. Therefore, in a stable state a lattice site is either empty or
	occupied by single particle of any one of the four colors. We have measured and plotted the densities $\rho_1(x,L)$
	and $\rho_3(x,L)$ of the color 1 and 3 particles in FIG. 6(a) for two different lattice sizes. Expectedly the $\rho_1(x,L)$
	and $\rho_3(x,L)$ curves are the mirror reflections of each other about the vertical line at $x/L=1/2$.
	In FIG. 6(b) we have plotted the total sum of these two densities $\rho_1(x,L) + \rho_3(x,L)$ against $x/L$.
	Both the curves have a flat region in the middle and the extent of this region gradually becomes larger as the
	lattice size increases. In FIG. 6(c) we have plotted $\rho(L)$ against $L^{-0.3655}$ to obtain the best fitted straight line.

\item [08.] {\bf Flush all version: Avalanche statistics.}

	In FIG. 7 we have shown three plots as follows:

	(a) The probability distribution $D(s,L)$ of avalanche sizes $s$ have been plotted for three different system sizes. 
	Each curve has two regimes, namely, a flat part for the small avalanche sizes which is followed by a decaying
	tail as usual for the avalanche size distributions. The presence of the horizontal part is quite unusual and we
	explain it in the following way.

	(b) The same data have been replotted using the scaled variables $D(s,L)L^{2.55}$ against $s/L^{1.7}$. The large
	avalanche size regime have been collapsed on top of one another nicely. This finite size scaling implies again
	the scaling exponents to be $\beta=2.55$ and $\alpha=1.7$ and therefore the avalanche size distribution exponent
	$\tau = \beta / \alpha = 1.5$, same as the original colored sandpile.

	(c) A second finite size scaling of the same data of (a) have been done when we plotted $D(s,L)L^{1.05}$ against
	$s / L$. This scaling gives a nice collapse of the data flat part and the peaks demarcating the small and big size
	avalanches.

        The horizontal part of the avalanche size distribution can be explained considering the fact that in the flush all
	version two or more particles of the same color meeting at any site create an avalanche where all these particles
	move together till they arrive at the boundary and then drop out from the system. If the origin of such a color 1 
	avalanche is at a distance $L-x$ from the boundary then the avalanche size is simply $L-x$. We have seen in figure 
	3(b) that $\rho_1(x,L)+\rho_2(x,L)$ is nearly flat in the middle. Therefore, if a color 1 particle is dropped on a 
	color 1 particle, or a color 3 particle is dropped on a color three particle, the avalanche sizes are $L-x$ or $x$. 
	Since the sum of the densities is constant, therefore such avalanches occur with uniform probabilities. Hence there 
	is the flat part in the probability distribution.

\begin{figure}[h]
	\includegraphics[width=17.0cm]{SuppFigure07.eps}
	\caption{Finite-size scaling plots have been shown for the probability distribution of the avalanche sizes of the 
	flush all version of the colored sandpile.}
\label{FIG07}
\end{figure}

\end {itemize}
}